\documentclass[a4paper,british]{scrartcl}
\usepackage[T1]{fontenc}
\usepackage[latin9]{inputenc}
\usepackage{amsmath}
\usepackage{graphicx}
\usepackage{amssymb}
\usepackage{esint}
\IfFileExists{url.sty}{\usepackage{url}}
                      {\newcommand{\url}{\texttt}}
\usepackage[numbers]{natbib}

\makeatletter
\newcommand{\lyxaddress}[1]{
\par {\raggedright #1
\vspace{1.4em}
\noindent\par}
}

\areaset{15cm}{23.8cm}
\renewcommand\@biblabel[1]{#1} 

\usepackage{babel}
\makeatother

\begin{document}

\title{A family tree of Markov models\\
in systems biology}

\author{{\normalsize Mukhtar Ullah\textsuperscript{a} and Olaf Wolkenhauer\textsuperscript{b}}}

\date{\small \today}

\maketitle

\lyxaddress{\noindent {\small Systems Biology and Bioinformatics Group, University
of Rostock, A.-Einstein-Str.~21, 18051 Rostock, Germany}\\
{\small \textsuperscript{a}Email: \url{mukhtar.ullah@uni-rostock.de}}\\
{\small \textsuperscript{b}Email: \url{olaf.wolkenhauer@uni-rostock.de},
Internet: \url{www.sbi.uni-rostock.de}, Tel/Fax: +49 381 498 75 70/72. }}

\begin{abstract}
Motivated by applications in systems biology, we seek a probabilistic
framework based on Markov processes to represent intracellular processes.
We review the formal relationships between different stochastic models
referred to in the systems biology literature. As part of this review,
we present a novel derivation of the differential Chapman-Kolmogorov
equation for a general multidimensional Markov process made up of
both continuous and jump processes. We start with the definition of
a time-derivative for a probability density but place no restrictions
on the probability distribution, in particular, we do not assume it
to be confined to a region that has a surface (on which the probability
is zero). In our derivation, the master equation gives the jump part
of the Markov process while the Fokker-Planck equation gives the continuous
part. We thereby sketch a {}``family tree'' for stochastic models
in systems biology, providing explicit derivations of their formal
relationship and clarifying assumptions involved.
\end{abstract}

\paragraph*{{\small Keywords:}}

{\small Markov processes, stochastic modelling, differential Chapman-Kolmogorov
equation, chemical master equations, Fokker-Planck equation, systems
biology.}{\small \par}

\section{Introduction}

Systems biology is a merger of systems theory with molecular and cell
biology. The key distinguishing feature of a systems biology approach
is the description of cell functions (e.g.~cell differentiation,
proliferation, apoptosis) as dynamic processes \citep{wolkenhauer:05,wolkenhauer:05b,wolkenhauer:05c}.
There are two dominant paradigms used in mathematical modelling of
biochemical reaction networks (pathways) in systems biology: the ``deterministic
approach'', using numerical simulations of nonlinear ordinary differential
equations (incl.~mass action type, power law or Michaelis-Menten
models), and the stochastic approach based on master equation and
stochastic simulations.

Key references in the area of stochastic modelling are the books by
\citet{kampen:92}, \citet{gillespie:92a}, \citet{breuer:02} and
\citet{gardiner:04}. Most stochastic models presented in these references
are derived on the basis of the Chapman-Kolmogorov equation (CKE),
a consistency condition on Markov processes, in the form of a system
of differential equations for the probability distribution. The system
of differential equations take the form of master equations for a
jump Markov process and Fokker-Planck equations (FPE) for a continuous
Markov process. For the way in which this happens, the reader is referred
to \citep{kampen:92} and \citep{gillespie:96,gillespie:92a}. For
a Markov process that is made up of both jump and continuous parts,
the differential equation takes the form of the differential Chapman-Kolmogorov
equation (dCKE) which has been derived in \citep{gardiner:04}. The
derivation is involved and requires the introduction of an arbitrary
function, which leads to boundary restrictions on the probability
distribution. As part of this review, we present a novel and more
concise derivation of the dCKE. Since most of the mathematical foundations
for stochastic models have been developed by physicists and mathematicians,
we hope that our derivation makes the theory more accessible to the
uninitiated researcher in the field of systems biology. We choose
Markov processes as a framework, since more realistic approaches for
modelling intracellular processes must take into account factors such
as heterogeneity of the environment, macromolecular crowding \citep{Schnell:04,Ellis:03}
and anomalous diffusion \citep{Saxton:94,Saxton:97,Golding:06}, to
name a few. Anomalous diffusion is described by fractional Fokker-Planck
equations \citep{Metzler:00}. Such treatments require advanced mathematical
formalisms which are beyond the level assumed in this paper.

The focus of the present paper is neither a comprehensive review of
stochastic approaches (See \citep{turner:04} for a recent survey,
\citep{paulsson:05,blomberg:06} for a recent theoretical analysis,
\citep{rao:02} for a discussion of the role of stochasticity in cell
biology) nor a comparison of the two approaches (e.g.~\citep{wolkenhauer:04,kummer:05,ullah:06}).
Instead, we review the formal relationships between the equations
referred to in the systems biology literature. We thereby, try to
establish a {}``family tree'' for stochastic models in systems biology,
providing explicit derivations of their formal relationship and clarifying
assumptions involved in a common framework (See Figure \ref{fig:stoch-eqs}).
 In the following section we focus on the origin of the chemical master
equation CME (a special form of the master equation for systems governed
by chemical reactions) within the framework of Markov processes. Such
generalisation provides a clearer picture of how the various stochastic
approaches used in systems biology are related within a common framework.

\section{Markov processes\label{sec:Markov-processes}}

Markov processes form the basis for the vast majority of stochastic
models of dynamical systems. The three books by \citet{gardiner:04},
\citet{kampen:92} and \citet{gillespie:92a} have become standard
references for the application of Markov processes to biological and
biochemical systems. At the centre of a stochastic analysis is the
so-called \emph{Chapman-Kolmogorov equation} (CKE) that describes
the evolution of a Markov process over time. From the CKE stem three
equations of practical importance: the \emph{master equation} for
jump-Markov processes, the \emph{Fokker-Planck equation} (FPE) for
continuous Markov processes and the \emph{differential Chapman-Kolmogorov
equation} (dCKE) for processes made up of both the continuous and
jump parts. A nice mathematical (but non-biological) account of these
equations can also be found in \citep{breuer:02}. Gardiner, van Kampen
and Gillespie take different approaches to derive these equations:

\begin{itemize}
\item Gillespie derives the FPE and the master equation independently from
the CKE and for the one-dimensional case only in \citep{gillespie:92a}.
In \citep{gillespie:92,gillespie:96} he extends the derivations to
multidimensional cases.
\item \citet{kampen:92} derives the master equation from the CKE for a
one-dimensional Markov process. The FPE is given as an approximation
of the master equation by approximating a jump processes with a continuous
one. The same approach is adopted in \citep{breuer:02}. However,
this should not mislead the reader to conclude that the FPE arises
this way. In fact, FPE is defined for a continuous Markov process.
\item \citet{gardiner:04} derives first the dCKE from the CKE for a multidimensional
Markov process whose probability distribution is assumed to be contained
in a closed surface. The FPE and the master equation are given as
special cases of the dCKE.
\end{itemize}
We start with a review of basic concepts from probability theory required
to read our proof. This is followed by a brief derivation of the CKE
and its graphical interpretation. From the CKE we derive the dCKE
and interpret its terms to show how the FPE and the master equation
appear as special cases of the dCKE. Finally we show that the CME
is just a special form of the master equation for jump processes governed
by chemical reactions.

A random variable $X$ describes a random event by assigning it values
$x$ (called \emph{states}) from a set $S$ (called \emph{state-spac}e)
and defines a probability distribution over this set. The set $S$
may be discrete, continuous or both. The probability distribution,
in the case of a discrete state-space $S=\left\{ n\right\} $, is
given by a set of probabilities $p_{n}$ such that\[
\mbox{Prob}\left\{ X=n\right\} =p_{n}\,.\]
In the case of a continuous state-space $S=\left\{ x\right\} $, the
probability distribution is given by a non-negative function $p(x)$
such that\[
\mbox{Prob}\left\{ x\leq X<x+ dx\right\} =p(x) dx\,.\]
In the literature, $p_{n}$ is referred to as a \emph{probability
mass function} (p.m.f) and $p(x)$ as a \emph{probability density
function} (p.d.f). The delta function $\delta(x)$ defined by\begin{equation}
\intop_{S} dx\, f(x)\delta(x-c)=f(c)\label{eq:delta}\end{equation}
for any function $f$ of $x$, allows us to write the discrete distribution
as a special case of the continuous case. Specifically we write\[
p(x)=\sum_{n}p_{n}\delta(x-n)\]
and note that \[
\mbox{Prob}\left\{ X=n\right\} =\intop_{x\leq n<x+ dx} dx\, p(x)=\intop_{x\leq n<x+ dx} dx\,\sum_{m}p_{m}\delta(x-m)=p_{n}\]
which is what we would expect in the discrete distribution. Since
the discrete distribution can always be derived easily from a continuous
one, we use hereafter the latter. When dealing with dynamical systems,
the probability distribution evolves over time. This leads to the
notion of a\emph{ stochastic process}, that is, a system in which
random variables are functions of time, written as $X(t)$. The states
are scalars for a one-dimensional system and vectors for a multidimensional
system. Note that, while the derivations of the master equation and
the Fokker-Planck equation in \citep{gillespie:92a,kampen:92} are
for the one-dimensional case only, we here present a general treatment
and derive all our results for multidimensional systems.

The probability distribution for an $N$-dimensional stochastic process\[
\mathbf{X}(t)=\left(X_{1}(t),\ldots,X_{N}(t)\right)\]
is written as\[
\textup{Prob}\left\{ \bigcap_{i=1}^{N}x_{i}\leq X_{i}(t)<x_{i}+ dx_{i}\right\} =p(x_{1},\ldots,x_{N},t) dx_{1}\cdots dx_{N}\,.\]
To simplify the notation, we use a short form

\[
\textup{Prob}\left\{ \mathbf{x}\leq\mathbf{X}(t)<\mathbf{x}+ d\mathbf{x}\right\} =p(\mathbf{x},t) d\mathbf{x}\,.\]
More useful will be the conditional probability density, $p(\mathbf{x},t\,|\,\mathbf{x}',t')$,
defined such that\[
\textup{Prob}\left\{ \mathbf{x}\leq\mathbf{X}(t)<\mathbf{x}+ d\mathbf{x}\,|\,\mathbf{X}(t')=\mathbf{x}'\right\} =p(\mathbf{x},t\,|\,\mathbf{x}',t') d\mathbf{x}\,.\]
When $t\geq t'$, $p(\mathbf{x},t\,|\,\mathbf{x}',t') d\mathbf{x}$
is called \emph{transition probability}. Since it is much easier to
work with densities $p(\cdot)$ rather than probabilities $\textup{Prob}\{\cdot\}$,
we shall use densities $p(\cdot)$, but abuse the terminology by referring
to it as ``probabilities''.

\begin{figure}[p]
\begin{centering}
\includegraphics{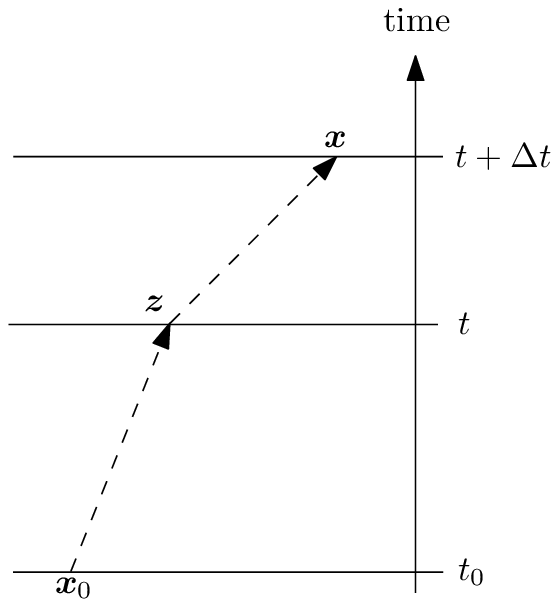}
\par\end{centering}

\caption{\label{fig:Chapman}Graphical interpretation of the Chapman-Kolmogorov
equation. The probability of a transition $\left(\mathbf{x}_{0}\rightarrow\mathbf{x}\right)$
can be obtained by summing probabilities of all transitions $\left(\mathbf{x}_{0}\rightarrow\mathbf{z}\rightarrow\mathbf{x}\right)$,
via intermediate states $\mathbf{z}$.}

\end{figure}

Essentially a Markov process is a stochastic process with a short
term memory. Mathematically it means that the conditional probability
of a state is determined entirely by the knowledge of the most recent
state. Specifically for any three successive times $t_{0}\leq t\leq t+\Delta t$
, one has\[
p(\mathbf{x},t+\Delta t\,|\,\mathbf{z},t;\mathbf{x}_{0},t_{0})=p(\mathbf{x},t+\Delta t\,|\,\mathbf{z},t)\]
where the conditional probability of $\mathbf{x}$ at $t+\Delta t$
is uniquely determined by the most recent state $\mathbf{z}$ at $t$
and is not affected by any knowledge of the initial state $\mathbf{x}_{0}$
at $t_{0}$. This Markov property is assumed to hold true for any
number of successive time intervals. To see how powerful this property
is, let us consider the factorisation of the joint probability \[
p(\mathbf{x},t+\Delta t\,;\,\mathbf{z},t)=p(\mathbf{x},t+\Delta t\,|\,\mathbf{z},t)p(\mathbf{z},t)\,.\]
Making both sides conditional on $(\mathbf{x}_{0},t_{0})$ will modify
this equation as \[
p(\mathbf{x},t+\Delta t\,;\,\mathbf{z},t\,|\,\mathbf{x}_{0},t_{0})=p(\mathbf{x},t+\Delta t\,|\,\mathbf{z},t;\mathbf{x}_{0},t_{0})p(\mathbf{z},t\,|\,\mathbf{x}_{0},t_{0})\]
which, by using the Markov property, reduces to\begin{equation}
p(\mathbf{x},t+\Delta t\,;\,\mathbf{z},t\,|\,\mathbf{x}_{0},t_{0})=p(\mathbf{x},t+\Delta t\,|\,\mathbf{z},t)p(\mathbf{z},t\,|\,\mathbf{x}_{0},t_{0})\,.\label{eq:joint_conditional}\end{equation}
The last equation shows that the joint probability can be expressed
in terms of transition probabilities. Recall the following rule for
joint probabilities\begin{equation}
p(x)=\int dy\, p(x,y)\label{eq:marginal}\end{equation}
which says that summing a joint probability over all values of one
of the variables eliminates that variable. Now integrating \eqref{eq:joint_conditional}
over $\mathbf{z}$ and using \eqref{eq:marginal}, we arrive at the
so-called \emph{Chapman-Kolmogorov equation} (CKE) \citep{gardiner:04}:
\begin{equation}
p(\mathbf{x},t+\Delta t\,|\,\mathbf{x}_{0},t_{0})=\int\limits _{S} d\mathbf{z}\, p(\mathbf{x},t+\Delta t|\mathbf{z},t)p(\mathbf{z},t|\mathbf{x}_{0},t_{0})\,.\label{eq:chapman}\end{equation}
This equation expresses the probability of a transition $\left(\mathbf{x}_{0}\rightarrow\mathbf{x}\right)$
as the summation of probabilities of all transitions $\left(\mathbf{x}_{0}\rightarrow\mathbf{z}\rightarrow\mathbf{x}\right)$
via intermediate states $\mathbf{z}$. Figure \ref{fig:Chapman} illustrates
the basic notion of a Markov process for which the CKE provides the
stochastic formalism. When the initial condition $(\mathbf{x}_{0},t_{0})$
is fixed, which is assumed here, the transition probability conditioned
on $(\mathbf{x}_{0},t_{0})$ is the same as the state probability:
\[
p(\mathbf{x},t)=p(\mathbf{x},t\,|\,\mathbf{x}_{0},t_{0})\,.\]

\section{Derivation of the dCKE}

The CKE serves as a description of a general Markov process, but cannot
be used to determine the temporal evolution of the probability. Here
we derive from the CKE a differential equation which will be more
useful in terms of describing the dynamics of the stochastic process.
Referred to as the \emph{differential} \emph{Chapman-Kolmogorov equation}
(dCKE) by \citet{gardiner:04}, this equation contains the CME as
a special case. This derivation is for a multidimensional Markov process.
We start with the definition of a time-derivative for a probability
density but place no restrictions on the probability distribution.
\citet{gardiner:04} instead starts with the expectation of an arbitrary
function which results in integration by parts and consequently the
need to assume that the probability density vanishes on the surface
of a region to which the process is confined. We do not need such
as assumption because of the simplicity of our approach. The master
equation gives the jump part of the Markov process while the Fokker-Planck
equation gives the continuous part.

Consider the time-derivative of the transition probability\begin{equation}
\frac{\partial}{\partial t}p(\mathbf{x},t)=\lim_{\Delta t\rightarrow0}\frac{1}{\Delta t}\Bigl\{ p(\mathbf{x},t+\Delta t)-p(\mathbf{x},t)\Bigr\}\,,\label{eq:diffp}\end{equation}
where differentiability of the transition probability with respect
to time is assumed. Employing the CKE \eqref{eq:chapman} and the
normalisation condition\[
\intop_{S} d\mathbf{z}\, p(\mathbf{z},t+\Delta t\,|\,\mathbf{x},t)=1\,,\]
since $p(\mathbf{z},t+\Delta t\,|\,\mathbf{x},t)$ is a probability,
\eqref{eq:diffp} can be rewritten as\[
\frac{\partial}{\partial t}p(\mathbf{x},t)=\lim_{\Delta t\rightarrow0}\frac{1}{\Delta t}\intop_{S} d\mathbf{z}\,\Biggl\{ p(\mathbf{x},t+\Delta t\,|\,\mathbf{z},t)p(\mathbf{z},t)-p(\mathbf{z},t+\Delta t\,|\,\mathbf{x},t)p(\mathbf{x},t)\Biggr\}\,.\]
Let us divide the region $S$ of integration into two regions based
on an arbitrarily small parameter $\epsilon>0$. The first region
$\left\Vert \mathbf{x}-\mathbf{z}\right\Vert <\epsilon$ corresponds
to a continuous state process and the above derivative in this region
will be denote by $I_{1}$. Here $\left\Vert \cdot\right\Vert $ denotes
a suitable vector-norm. The second region $\left\Vert \mathbf{x}-\mathbf{z}\right\Vert \geq\epsilon$
corresponds to a jump process and the above derivative in this region
will be denote by $I_{2}$. Since \eqref{eq:diffp-2} gives the derivative
in the whole region $S$, we can write\begin{equation}
\frac{\partial}{\partial t}p(\mathbf{x},t)=I_{1}+I_{2},\label{eq:diffp-2}\end{equation}
where\[
I_{1}=\lim_{\Delta t\rightarrow0}\frac{1}{\Delta t}\intop_{\left\Vert \mathbf{x}-\mathbf{z}\right\Vert <\epsilon} d\mathbf{z}\,\Biggl\{ p(\mathbf{x},t+\Delta t\,|\,\mathbf{z},t)p(\mathbf{z},t)-p(\mathbf{z},t+\Delta t\,|\,\mathbf{x},t)p(\mathbf{x},t)\Biggr\},\]
and\[
I_{2}=\lim_{\Delta t\rightarrow0}\frac{1}{\Delta t}\intop_{\left\Vert \mathbf{x}-\mathbf{z}\right\Vert \ge\epsilon} d\mathbf{z}\,\Biggl\{ p(\mathbf{x},t+\Delta t\,|\,\mathbf{z},t)p(\mathbf{z},t)-p(\mathbf{z},t+\Delta t\,|\,\mathbf{x},t)p(\mathbf{x},t)\Biggr\}\,.\]
In the first region $\left\Vert \mathbf{x}-\mathbf{z}\right\Vert <\epsilon$,
the integrand of $I_{1}$ can be expanded in powers of $\mathbf{x}-\mathbf{z}$
using a Taylor expansion. Setting $\mathbf{x}-\mathbf{z}=\mathbf{r}$,
we can write,\begin{equation}
I_{1}=\lim_{\Delta t\rightarrow0}\frac{1}{\Delta t}\intop_{\left\Vert \mathbf{r}\right\Vert <\epsilon} d\mathbf{r}\,\Biggl\{ p(\mathbf{x},t+\Delta t\,|\,\mathbf{x}-\mathbf{r},t)p(\mathbf{x}-\mathbf{r},t)-p(\mathbf{x}-\mathbf{r},t+\Delta t\,|\,\mathbf{x},t)p(\mathbf{x},t)\Biggr\}\,.\label{eq:I1}\end{equation}
In order to expand the integrand more easily into a Taylor series,
let us define a function\[
f(\mathbf{x};\mathbf{r})\triangleq p(\mathbf{x}+\mathbf{r},t+\Delta t\,|\,\mathbf{x},t)p(\mathbf{x},t)\]
so that the integrand in \eqref{eq:I1} becomes $f(\mathbf{x}-\mathbf{r};\mathbf{r})-f(\mathbf{x};-\mathbf{r})$,
which, after a Taylor expansion, becomes\[
-f(\mathbf{x};-\mathbf{r})+f(\mathbf{x};\mathbf{r})+\sum_{i}(-r_{i})\frac{\partial f(\mathbf{x};\mathbf{r})}{\partial x_{i}}+\frac{1}{2}\sum_{i,j}r_{i}r_{j}\frac{\partial^{2}f(\mathbf{x};\mathbf{r})}{\partial x_{i}\partial x_{j}}+\mbox{higher-order terms\,.}\]
The integrals of the first two terms cancel because of the symmetry
$\int f(\mathbf{x};\mathbf{r}) d\mathbf{r}=\int f(\mathbf{x};-\mathbf{r}) d\mathbf{r}$,
when the integral is over all the positive and negative values of
$\mathbf{r}$ in the region. Thus, we have

\begin{multline*}
I_{1}=\lim_{\Delta t\rightarrow0}\frac{1}{\Delta t}\intop_{\left\Vert \mathbf{r}\right\Vert <\epsilon} d\mathbf{r}\,\Biggl\{-\sum_{i}r_{i}\frac{\partial}{\partial x_{i}}\Bigl[p(\mathbf{x}+\mathbf{r},t+\Delta t\,|\,\mathbf{x},t)p(\mathbf{x},t)\Bigr]\\
+\frac{1}{2}\sum_{i,j}r_{i}r_{j}\frac{\partial^{2}}{\partial x_{i}\partial x_{j}}\Bigl[p(\mathbf{x}+\mathbf{r},t+\Delta t\,|\,\mathbf{x},t)p(\mathbf{x},t)\Bigr]+\mbox{higher-order terms}\Biggr\}\,.\end{multline*}
For the state increments $X_{i}(t+\Delta t)-X_{i}(t)$, recognising
the (conditional) expectations \[
\langle X_{i}(t+\Delta t)-X_{i}(t)\,|\,\mathbf{X}(t)=\mathbf{x}\rangle=\intop_{\left\Vert \mathbf{r}\right\Vert <\epsilon} d\mathbf{r}\, r_{i}\, p(\mathbf{x}+\mathbf{r},t+\Delta t\,|\,\mathbf{x},t)\]
and\begin{multline*}
\langle\left[X_{i}(t+\Delta t)-X_{i}(t)\right]\left[X_{j}(t+\Delta t)-X_{j}(t)\right]\,|\,\mathbf{X}(t)=\mathbf{x}\rangle\\
=\intop_{\left\Vert \mathbf{r}\right\Vert <\epsilon} d\mathbf{r}\, r_{i}r_{j}\, p(\mathbf{x}+\mathbf{r},t+\Delta t\,|\,\mathbf{x},t),\end{multline*}
we refer to the differentiability conditions for continuous processes,
i.e., $\left\Vert \mathbf{x}-\mathbf{z}\right\Vert <\epsilon$ \citep[section 3.4]{gardiner:04}:\begin{align}
\lim_{\Delta t\rightarrow0}\frac{\langle X_{i}(t+\Delta t)-X_{i}(t)\,|\,\mathbf{X}(t)=\mathbf{x}\rangle}{\Delta t} & =A_{i}(\mathbf{x},t)+o(\epsilon)\label{eq:DriftCoeff}\\
\lim_{\Delta t\rightarrow0}\frac{\langle\left[X_{i}(t+\Delta t)-X_{i}(t)\right]\left[X_{j}(t+\Delta t)-X_{j}(t)\right]\,|\,\mathbf{X}(t)=\mathbf{x}\rangle}{\Delta t} & =B_{ij}(\mathbf{x},t)+o(\epsilon)\label{eq:DiffusionCoeff}\end{align}
where $o(\epsilon)$ represents vanishing terms, such that $\lim_{\epsilon\rightarrow0}o(\epsilon)/\epsilon=0$.
The higher-order terms involve higher-order coefficients which must
vanish. To see that, for the third-order coefficient \[
\lim_{\Delta t\rightarrow0}\frac{1}{\Delta t}\intop_{\left\Vert \mathbf{r}\right\Vert <\epsilon} d\mathbf{r}\, r_{i}r_{j}r_{k}\, p(\mathbf{x}+\mathbf{r},t+\Delta t\,|\,\mathbf{x},t)=C_{ijk}(\mathbf{x},t)+o(\epsilon)\,.\]
However\begin{eqnarray*}
\lefteqn{\lim_{\Delta t\rightarrow0}\frac{1}{\Delta t}\intop_{\left\Vert \mathbf{r}\right\Vert <\epsilon} d\mathbf{r}\, r_{i}r_{j}r_{k}\, p(\mathbf{x}+\mathbf{r},t+\Delta t\,|\,\mathbf{x},t)}\\
 & \leq & \left\Vert \mathbf{r}\right\Vert \lim_{\Delta t\rightarrow0}\frac{1}{\Delta t}\intop_{\left\Vert \mathbf{r}\right\Vert <\epsilon} d\mathbf{r}\, r_{i}r_{j}\, p(\mathbf{x}+\mathbf{r},t+\Delta t\,|\,\mathbf{x},t)\\
 & \leq & \epsilon\left[B_{ij}(\mathbf{x},t)+o(\epsilon)\right]\\
 & \leq & o(\epsilon)\,.\end{eqnarray*}
Hence $C(\mathbf{x},t)$ must vanish. The vanishing of higher-order
coefficients follows immediately. In physics, the vector $A(\mathbf{x},t)$
is known as the ``\emph{drift vector}'' and the matrix $B(\mathbf{x},t)$
as the ``\emph{diffusion matrix}''. This terminology is suggested
by the observation that, given $X(t)=x$, the state increment vector
$\mathbf{X}(t+ dt)-\mathbf{X}(t)$ for a continuous process has
a mean approaching $A(\mathbf{x},t) dt$ and a covariance approaching
$B(\mathbf{x},t) dt$, as $\epsilon$ approaches zero. This also
suggests the following update rule, for $\epsilon\rightarrow0$ and
under assumptions given in \citep[section 3.5.2]{gardiner:04}, \begin{equation}
\mathbf{X}(t+ dt)=\mathbf{X}(t)+A(\mathbf{X}(t),t) dt+\left[B(\mathbf{X}(t),t) dt\right]^{1/2}\label{eq:langevin}\end{equation}
which is a form of the {}``\emph{Langevin equation}'' \citep{kampen:92}.
We remark here that \eqref{eq:DriftCoeff} and \eqref{eq:DiffusionCoeff}
are postulated here for mathematical convenience. A more rigorous
justification is given in \citep{gillespie:96}.  Subject to the differentiability
conditions \eqref{eq:DriftCoeff} and \eqref{eq:DiffusionCoeff},
we see that as $\epsilon\rightarrow0$, \begin{equation}
I_{1}\rightarrow-\sum_{i}\frac{\partial}{\partial x_{i}}\Bigl[A_{i}(\mathbf{x},t)p(\mathbf{x},t)\Bigr]+\frac{1}{2}\sum_{i,j}\frac{\partial^{2}}{\partial x_{i}\partial x_{j}}\Bigl[B_{ij}(\mathbf{x},t)p(\mathbf{x},t)\Bigr]\,.\label{eq:I1f}\end{equation}
Next we work out the jump probability rate \[
I_{2}=\lim_{\Delta t\rightarrow0}\frac{1}{\Delta t}\intop_{\left\Vert \mathbf{x}-\mathbf{z}\right\Vert \geq\epsilon} d\mathbf{z}\,\Biggl\{ p(\mathbf{x},t+\Delta t\,|\,\mathbf{z},t)p(\mathbf{z},t)-p(\mathbf{z},t+\Delta t\,|\,\mathbf{x},t)p(\mathbf{x},t)\Biggr\}\,.\]
We will use the differentiability condition for jump processes, i.e.,
$\left\Vert \mathbf{x}-\mathbf{z}\right\Vert \geq\epsilon$ \citep[section 3.4]{gardiner:04}:\[
\lim_{\Delta t\rightarrow0}\frac{1}{\Delta t}p(\mathbf{x},t+\Delta t\,|\,\mathbf{z},t)=W(\mathbf{x}\,|\,\mathbf{z},t)\,,\]
where $W(\mathbf{x}\,|\,\mathbf{z},t)$ is called the transition rate
for the jump $(\mathbf{z}\rightarrow\mathbf{x})$. Subject to this
condition, we see that as $\epsilon\rightarrow0$, the region of integration
approaches $S$, leading to\begin{equation}
I_{2}\rightarrow\intop_{S} d\mathbf{z}\,\Bigl[W(\mathbf{x}\,|\,\mathbf{z},t)p(\mathbf{z},t)-W(\mathbf{z}\,|\,\mathbf{x},t)p(\mathbf{x},t)\Bigr]\,.\label{eq:I2f}\end{equation}
Adding \eqref{eq:I1f} and \eqref{eq:I2f}, we can rewrite \eqref{eq:diffp-2}
to arrive at the dCKE:\begin{multline}
\frac{\partial}{\partial t}p(\mathbf{x},t)=-\sum_{i}\frac{\partial}{\partial x_{i}}\Bigl[A_{i}(\mathbf{x},t)p(\mathbf{x},t)\Bigr]+\frac{1}{2}\sum_{i,j}\frac{\partial^{2}}{\partial x_{i}\partial x_{j}}\Bigl[B_{ij}(\mathbf{x},t)p(\mathbf{x},t)\Bigr]\\
+\intop_{S} d\mathbf{z}\,\Bigl[W(\mathbf{x}\,|\,\mathbf{z},t)p(\mathbf{z},t)-W(\mathbf{z}\,|\,\mathbf{x},t)p(\mathbf{x},t)\Bigr]\,\label{eq:diff-chapman}\end{multline}
We now have a differential equation characterising the dynamics of
the probability distribution $p(\mathbf{x},t)$, that is the probability
of a state at any time, starting from a given initial probability
distribution. This completes our derivation of the differential Chapman-Kolmogorov
equation. Differences between our derivation and those available in
the literature are described in Section \ref{sec:Conclusions} (Conclusions).
The following section will classify Markov processes based on this
dCKE. This is followed by a derivation of the chemical master equation.
Finally, in Section \ref{sec:CME-in-SB}, we discuss the use of the
master equation in systems biology.

\section{Classification of Markov processes based on the dCKE\label{sec:Interpret-CKE} }

Being a linear differential equation, the dCKE is more convenient
for mathematical treatment than the original CKE. More importantly,
it has a more direct physical interpretation. The coefficients $A(\mathbf{x},t)$,
$B(\mathbf{x},t)$ and $W(\mathbf{z}\,|\,\mathbf{x},t)$ are specified
by the system under consideration, and thus the solution of the dCKE
gives the probability distribution for the state of the given system
\citep{kampen:92}. The original CKE, on the other hand, has no specific
information about any particular Markov process. We now interpret
the different terms of \eqref{eq:diff-chapman}. Following \citep{gardiner:04,breuer:02}
we first consider the case\[
B_{ij}(\mathbf{x},t)=W(\mathbf{x}\,|\,\mathbf{z},t)=W(\mathbf{z}\,|\,\mathbf{x},t)=0,\]
reducing the dCKE to \[
\frac{\partial}{\partial t}p(\mathbf{x},t)=-\sum_{i}\frac{\partial}{\partial x_{i}}\Bigl[A_{i}(\mathbf{x},t)p(\mathbf{x},t)\Bigr]\,,\]
which is a special case of the so-called \emph{``Liouville equation'',}
describing a deterministic motion (See \citep[section 3.5.3]{gardiner:04}):\[
\frac{ d}{ dt}\mathbf{x}(t)=A(\mathbf{x},t)\,.\]
This is the simplest example of a Markov process.

Next, if $A(\mathbf{x},t)=B(\mathbf{x},t)=0$, the CKE reduces to
\begin{equation}
\frac{\partial}{\partial t}p(\mathbf{x},t)=\intop_{S} d\mathbf{z}\,\Bigl[W(\mathbf{x}\,|\,\mathbf{z},t)p(\mathbf{z},t)-W(\mathbf{z}\,|\,\mathbf{x},t)p(\mathbf{x},t)\Bigr]\,.\label{eq:master}\end{equation}
This is called the \emph{``master equation''} describing jump-Markov
process with discontinuous sample paths.

Next, if $W(\mathbf{x}\,|\,\mathbf{z},t)=W(\mathbf{z}\,|\,\mathbf{x},t)=0$,
the CKE reduces to \[
\frac{\partial}{\partial t}p(\mathbf{x},t)=-\sum_{i}\frac{\partial}{\partial x_{i}}\Bigl[A_{i}(\mathbf{x},t)p(\mathbf{x},t)\Bigr]+\frac{1}{2}\sum_{i,j}\frac{\partial^{2}}{\partial x_{i}\partial x_{j}}\Bigl[B_{ij}(\mathbf{x},t)p(\mathbf{x},t)\Bigr]\,,\]
which is called the \emph{``Fokker-Planck equation''} (FPE) and is
equivalent to the Langevin's equation \eqref{eq:langevin} under the
conditions given in \citep{kampen:92,gillespie:96,gardiner:04}. The
corresponding process is known as a \emph{diffusion process} which
is continuous but not deterministic. This shows that the FPE is originally
defined for a continuous process. However the FPE can also arise as
an approximation of the master equation when the jumps of the corresponding
discrete process are assumed to be small \citep{kampen:92,Kepler:01}.

Finally we consider the case where the diffusion matrix $B(\mathbf{x},t)=0$,
which leads us to\[
\frac{\partial}{\partial t}p(\mathbf{x},t)=-\sum_{i}\frac{\partial}{\partial x_{i}}\Bigl[A_{i}(\mathbf{x},t)p(\mathbf{x},t)\Bigr]+\intop_{S} d\mathbf{z}\,\Bigl[W(\mathbf{x}\,|\,\mathbf{z},t)p(\mathbf{z},t)-W(\mathbf{z}\,|\,\mathbf{x},t)p(\mathbf{x},t)\Bigr]\,.\]
which is called the \emph{``Liouville master equation''} (LME) in
\citep[chap. 1]{breuer:02} and describes a piecewise deterministic
process with sample paths consisting of smooth deterministic pieces
interrupted by instantaneous jumps. One way in which the LME arises
is when an originally jump Markov process is approximated by a process
with discrete and continuous components \citep{lipniacki:06,Paszek:07}. 

In the most general case, when none of the quantities $A(\mathbf{x},t)$,
$B(\mathbf{x},t)$ and $W(\mathbf{z}\,|\,\mathbf{x},t)$ vanish, the
dCKE may describe a process whose sample paths are piecewise continuous,
made up of pieces which correspond to a diffusion process with a nonzero
drift, onto which is superimposed a fluctuating part.

\section{The chemical master equation}

\begin{figure}[p]
\begin{centering}
\includegraphics{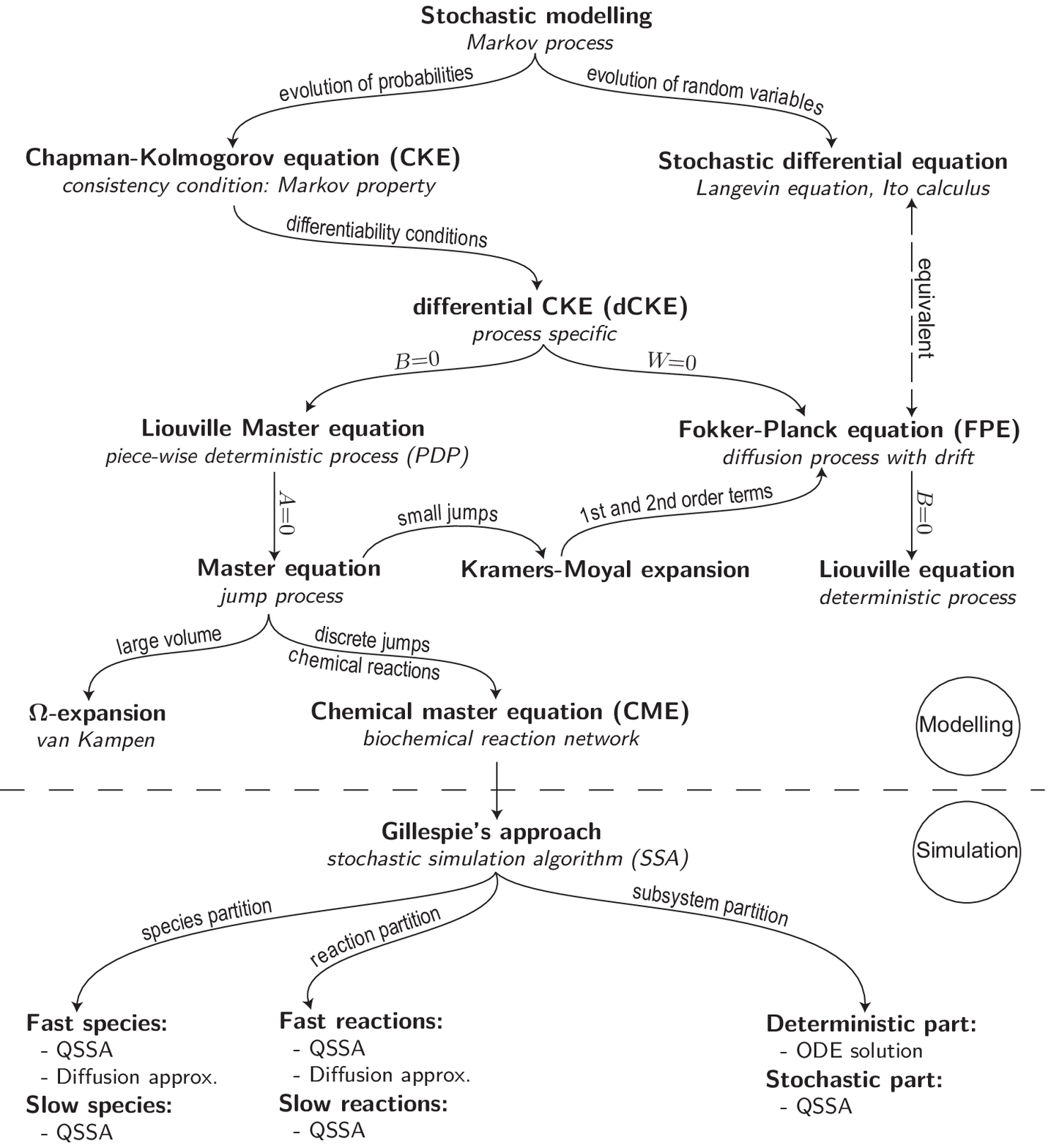}
\par\end{centering}

\caption{Interrelationships for the stochastic models and their simulation
which are covered in this paper. The coefficients $A,B,W$ respectively
refer to the drift-vector, diffusion-matrix and the transition-rate
in the dCKE. QSSA stands for Quasi-Steady-State Assumption.\label{fig:stoch-eqs}}

\end{figure}

Consider a Markov process with a discrete state-space. The master
equation for this discrete process can be obtained from \eqref{eq:master}
to give \[
\frac{\partial}{\partial t}p_{\mathbf{n}}(t)=\sum_{\mathbf{m}}\Bigl[W(\mathbf{n}\,|\,\mathbf{m},t)p_{\mathbf{m}}(t)-W(\mathbf{m}\,|\,\mathbf{n},t)p_{\mathbf{n}}(t)\Bigr],\]
where $\mathbf{m}$ the intermediate state, and $\mathbf{n}$ the
final state. Since $p_{\mathbf{n}}(t)$ is a probability (and not
a density), the integral $\int_{S}$ is replaced with the summation
$\sum_{\mathbf{m}}$. We can rewrite this equation in terms of jumps
$\mathbf{r}=\mathbf{n}-\mathbf{m}$,\begin{equation}
\frac{\partial}{\partial t}p_{\mathbf{n}}(t)=\sum_{\mathbf{r}}\Bigl[W(\mathbf{n}\,|\,\mathbf{n}-\mathbf{r},t)p_{\mathbf{n}-\mathbf{r}}(t)-W(\mathbf{n}+\mathbf{r}\,|\,\mathbf{n},t)p_{\mathbf{n}}(t)\Bigr]\,,\label{eq:master-discrete}\end{equation}
where we have used the symmetry $\sum_{\mathbf{r}}\phi(\mathbf{-r})=\sum_{\mathbf{r}}\phi(\mathbf{r})$,
for an arbitrary function $\phi(\cdot)$, when writing the second
summand. Now consider an $N$-component and $M$-reaction biochemical
system. Let $i$ label the different components (chemical species)
and $j$ label different reaction channels. The copy number of $i$th
component at the variable time $t$ will be denoted by $X_{i}(t)$
which takes values $n_{i}$ from the set of whole numbers. Each occurrence
of $j$-th reaction channel changes the copy number $n_{i}$ of $i$th
component by an amount $\nu_{ij}$. The elements $\nu_{ij}$ form
the stochiometric matrix $\nu$ whose $j$th column will be denoted
by $\nu_{j}$. It is assumed that the species are distributed homogeneously
(well mixed) in a closed system of constant volume $\Omega$ at a
constant temperature. This essentially assumes that changes only depend
on the current state (Markov property) and that we can avoid spatial
considerations \citep{kampen:92,gardiner:04,elf:01} and macromolecular
crowding \citep{grima:06}. However, since diffusion may not always
be rapid, spatial considerations become important when dealing with
intracellular processes \citep{elf:04,kruse:06,kholodenko:06}. Here
we are interested in a stochastic formulation which dates back to
the initial work by \citet{kramers:40}. Under the stated assumptions,
the vector $\mathbf{X}(t)=(X_{1}(t),\ldots,X_{N}(t))$ taking values
$\mathbf{n}=(n_{1},\ldots,n_{N})$ is a continuous time Markov process.
The jump sizes $\mathbf{\nu}$ are determined by the stoichiometry
and molecularity of the reactions and, therefore, can only take values
from the set $\left\{ \mathbf{\nu}_{1},\ldots,\mathbf{\nu}_{M}\right\} $
of the elementary changes. Thus, for our system of chemical reactions,
\eqref{eq:master-discrete} becomes\[
\frac{\partial}{\partial t}p_{\mathbf{n}}(t)=\sum_{j=1}^{M}\Bigl[W(\mathbf{n}\,|\,\mathbf{n}-\mathbf{\nu}_{j},t)p_{\mathbf{n}-\mathbf{\nu}_{j}}(t)-W(\mathbf{n}+\mathbf{\nu}_{j}\,|\,\mathbf{n},t)p_{\mathbf{n}}(t)\Bigr]\,.\]
Since $\mathbf{\nu}_{j}$ is uniquely defined for a reaction $R_{j}$,
we introduce a simpler notation\[
a_{j}(\mathbf{n})\triangleq W(\mathbf{n}+\mathbf{\nu}_{j}\,|\,\mathbf{n},t)\]
to rewrite the above master equation as: \begin{equation}
\dfrac{\partial}{\partial t}p_{\mathbf{n}}(t)=\sum_{j=1}^{M}\left[a_{j}(\mathbf{n}-\mathbf{\nu}_{j})p_{\mathbf{n}-\mathbf{\nu}_{j}}(t)-a_{j}(\mathbf{n})p_{\mathbf{n}}(t)\right],\label{eq:cme}\end{equation}
which is referred as the chemical master equation in the systems biology
literature \citep{singer:53,gillespie:77,kampen:92}. This shows that
the CME is just a special form of the master equation for jump processes
governed by chemical reactions. The coefficient $a_{j}$ is referred
to as the \emph{reaction propensity} and is interpreted such that
$a_{j}(\mathbf{n}) dt$ gives the probability of $j$th reaction
occurring in the time interval $[t,t+ dt)$ from state $\mathbf{n}$
at time $t$. In the stochastic setting of \citet{gillespie:77},
the $j$th reaction channel is characterised by a \emph{stochastic
rate constant} $c_{j}$ such that $c_{j} dt$ gives the probability
that a particular combination of molecules will react according the
$j$th channel in the next infinitesimal interval $\left[t,t+ dt\right]$.
The propensity $a_{j}(n)$ is thus $c_{j}$ times the number $h_{j}(n)$
of different possible ways in which molecules can combine to react
according the $j$th channel. Since this equation is difficult to
solve analytically or even numerically, several attempts have been
made to avoid a direct solution or simulation of the CME. The most
successful implementation is the stochastic simulation algorithm (SSA)
which originated from work by \citet{doob:45} but it was Gillespie
who pioneered its use for generating sample paths of chemical reaction
networks \citep{gillespie:76,gillespie:77}. The SSA is widely used
in systems biology \citep{dublanche:06,lipniacki:06,lipniacki:06a,paszek:05,paulsson:05}.
Figure \ref{fig:stoch-eqs} provides an overview of stochastic models
and interrelationships referred to here. %

This completes our formal analysis and we now return to application
of the CME in systems biology.

\section{Master equations in systems biology\label{sec:CME-in-SB}}

The chemical master equation \eqref{eq:cme} is the basis for most
stochastic models in systems biology. For complex systems, involving
large numbers of chemical species and reactions, the computational
cost may be considerable. For this reason modifications to the algorithm
and strategies to simplify the model prior to a computer simulation
have been suggested. The efforts to reduce the computational complexity
of stochastic approaches can be grouped into approximate stochastic
methods and hybrid methods. Approximate stochastic methods try to
speed up the simulation by compromising exactness of the master equation
whereas the hybrid methods treat parts of the system deterministically
and other parts stochastically. What follows is a brief discussion
of the systems biology literature, using the CME and SSA, including
strategies that have been developed to reduce the computational complexity.

\subsection{Approximate stochastic methods}

In \citep{gillespie:01} Gillespie presents an approximate and thereby
faster simulation method known as the $\tau$-leap method. Instead
of simulating individual reactions, the number of reactions of each
type in a sequence of short time intervals is simulated. Optimal ways
to select the leap-length of the intervals have been investigated
in \citep{gillespie:03,cao:06}. In \citep{rao:03} Rao et al.~propose
a reduction of a chemical system by partitioning molecular species
into slow (primary) and fast (intermediate) molecular species. Assuming
that the intermediate species (conditional on the primary species)
are Markovian, they apply the \emph{quasi-steady-state assumption}
(QSSA) which essentially assumes that the conditional probability
distribution of the intermediate species is time-invariant (i.e.,~it
has reached a steady-state); thereby eliminating these species from
the chemical master equation. A modified version of Gillespie algorithm
is subsequently proposed to simulate the resulting reduced system
for slow species. In \citep{haseltine:02,haseltine:05} Haseltine
and coworker also use the partitioning method for reduction, but partition
chemical reactions into fast and slow reactions. The fast reactions
are approximated by using Langevin or deterministic equations. The
coefficients of the reduced chemical master equation are influenced
by the fast reactions. The authors propose simulation algorithms for
the slow reactions, subject to constraints imposed by fast reactions.
The idea of partitioning to speed up slow-scale simulation has also
been used in \citep{cao:05,cao:05a,cao:05b,goutsias:05}. \citet{salis:05}
present a probabilistic steady-state approximation that separates
the time scales of an arbitrary reaction network, detects the convergence
of a marginal distribution to a quasi-steady-state, directly samples
the underlying distribution, and uses those samples to predict the
state of the system. \citet{elf:03a} propose that, in case of higher
dimensions, the master equation could be approximated by the FPE and
then discretised in space and time by a finite difference method.
They demonstrate the method for a four-dimensional problem in the
regulation of cell processes and compare it to the Monte Carlo method
of Gillespie. \citet{paulsson:00a} use the CME to analyse a negative
feedback system composed of two species regulating the synthesis of
each other.  In \citep{paulsson:04} Paulsson uses a variant of the
fluctuation-dissipation theorem to give a generic expression for noise
arising from different cellular processes, applying the theorem to
a simple generic model representing simple gene expression. Paulsson
uses the notion of an $\Omega$-expansion of the master equation,
i.e., a Taylor series expansion in powers of $1/\Omega$, where $\Omega$
is a system size parameter \citep{kampen:92}. The first- and second-
order terms of the expansion reproduce the macroscopic rate equations
and realise the fluctuation-dissipation theorem respectively. An equation
that decomposes the intrinsic and extrinsic noise contributions is
derived to simplify the analysis. \citet{elf:03} give a general method
to simplify the master equation in a linear noise approximation (LNA),
obtained through an $\Omega$-expansion of the master equation. They
derive the LNA for the stationary state of a general system of chemical
reactions and use it to estimate sizes, correlations and time scales
of stochastic fluctuations. They demonstrate that the LNA allows a
rapid characterisation of stochastic properties for intracellular
networks over a large parameter space. They also show that the LNA
can be made more accurate in cases where fast variables can be eliminated
from the system. In \citep{elf:03a} the same authors use the LNA
of the master equation to characterise intracellular metabolite fluctuations.
In both publications, the results of the LNA are compared to simulation
of the master equation.

\subsection{Hybrid methods}

\citet{alfonsi:05} present a simulation approach for hybrid stochastic
and deterministic reaction models. The system is adaptively partitioned
into deterministic and stochastic parts based on a given criteria
at each time step of the simulation. They present two algorithmic
schemes. The `direct hybrid method' explicitly calculates which reaction
occurs and when it occurs. The `first and next reaction method' generates
a putative time for each reaction. The reaction corresponding to the
smallest time is chosen to occur; the state according updated and
the process repeated. \citet{lipniacki:06} derive the first-order
partial differential equations for probability distribution function
from stochastic differential equations describing approximate kinetics
of a single cell. Resulting equations are used to calculate mRNA-protein
distribution in the case of single gene regulation and the protein-protein
distribution in the case of two-gene regulatory systems. In \citep{lipniacki:06a}
the same authors present a hybrid stochastic and deterministic treatment
of the NF-$\kappa$B regulatory module to analyse a single cell regulation.
They combine ordinary differential equations, used for the description
of fast reaction channels of processes involving a large number of
molecules, with a stochastic switch to account for the activity of
the genes involved. \citet{Paszek:07} makes two approximations to
the exact stochastic description of a gene regulatory network: the
continuous approximation considering only the stochasticity due to
the gene activity; and the mixed approximation attributing the additional
stochasticity to the mRNA transcription/decay process. The underlying
distribution is then described by a system of partial differential
equations derived from the dCKE after specific assumptions on the
coefficients $A,B$ and $W$. \citet{kummer:05} take a different
approach by deciding on the fly which approach to use. Studying the
stochastic simulation of signal transduction via calcium, the authors
observe that the transition from stochastic to deterministic occurs
within a range of particle numbers that depends the phase space of
the system.

Yet another approach is given in \citep{ochiai:05} where a novel
multidimensional stochastic framework is proposed to model multi-gene
expression dynamics. Inspired by the dCKE, they propose a new experimental
scheme which will measure the instantaneous transition probabilities.
Given experimental data, one could obtain the coefficients of the
dCKE, which can then be solved to obtain the distribution and the
corresponding moments and correlations. \citet{dublanche:06} analyse
noise in a negatively feedback-regulated transcription factor (TF)
and the effects of the feedback loop on a gene repressed by the same
TF using a modified Gillespie algorithm (provided by SmartCell). The
authors find that within a certain range of repression strength, the
negative feedback loop minimises the noise whereas outside this range,
noise is increased. It is proposed that this may arise from plasmid
fluctuations.

\section{Conclusions\label{sec:Conclusions}}

The application of stochastic models is usually motivated by uncertainty
arising from variability. In the engineering and physical sciences
the variability arises mainly from measurements. In systems biology
the variability arises mainly from the complexity of intracellular
processes. By complexity we mean the fact that in a particular biological
network we are forced to eliminate many variables that are influencing
the observation we make. While the assumptions of constant temperature,
pH level, volume and water balance may not worry most modellers, the
large number of unmodelled variables may be of greater concern. Implicit
in most stochastic models is the assumption of a well-mixed homogeneous
environment in which there are more non-reactive collisions than the
reactive ones. More realistic spatial representations are therefore
an increasingly important research theme in systems biology. In gene
expression a very small number of molecules controls potentially very
large molecular populations, suggesting hybrid approaches to combine
stochastic and deterministic formalisms. Many cellular processes,
for instance the differentiation of stem cells, are multistable systems
for which state-of-the-art single-cell measurements are providing
increasingly valuable data with nanometre and millisecond resolution.
The advance of these technologies allows us to monitor the transcription
of individual genes, leading also to a demand for advanced stochastic
modelling and simulation.

Motivated by applications of stochastic models in systems biology,
we described a probabilistic framework based on Markov processes to
represent biochemical reaction networks. We provided a novel derivation
of the differential Chapman-Kolmogorov equation for a general Markov
process made up of both continuous and jump processes. Then we reviewed
the formal relationships between the equations referred to in the
systems biology literature, establishing a {}``family tree'' for
stochastic models in systems biology, providing explicit derivations
of their formal relationship and clarifying assumptions involved in
a common framework (See Figure \ref{fig:stoch-eqs}). Our derivation
starts with the definition of a time-derivative unlike Gardiner who
starts with the expectation of an arbitrary function. We place no
restrictions on the probability distribution, whereas Gardiner assumes
it to be confined to a region that has a surface, and the probability
being zero on the surface. The master equation gives the jump part
of the Markov process while the FPE gives the continuous part. The
derivation of FPE in \citet{breuer:02} and \citet{kampen:92} involves
approximation of the master equation by investigating the limit of
a jump process.

\section*{Acknowledgements}

O.W. and M.U. acknowledge the support of the German Federal Ministry
of Education and Research (BMBF) grant 01GR0475 as part of the National
Genome Research Network (NGFN-SMP Protein). Further support was provided
by grants from the European Commissions FP6, projects AMPKIN and COSBICS
and the UK Department for the Environment, Food and Rural Affairs
(DEFRA). Discussions with Allan Muir, Johan Elf provided valuable
help in preparing the manuscript.

\bibliographystyle{plainnat}
\bibliography{ref_ullah}

\end{document}